\documentclass[aps,floats,twocolumn,epsf,prl,showpacs]{revtex4-1}
\usepackage{graphicx}
\usepackage{epstopdf}
\usepackage[colorlinks=true,urlcolor=blue,citecolor=blue,linkcolor=blue,breaklinks=true]{hyperref}
\usepackage{color}

\begin{document}

\title{Quantum criticality in the two-dimensional periodic Anderson model}

\author{T. Sch\"afer$^{a,b,c}$, A. A. Katanin$^d$, M. Kitatani$^a$, A. Toschi$^a$ and K. Held$^a$}

\affiliation{$^a$Institute of Solid State Physics, TU Wien, 1040 Vienna, Austria}
\affiliation{$^b$Coll{\`e}ge de France, 11 place Marcelin Berthelot, 75005 Paris, France}
\affiliation{$^c$CPHT, CNRS, Ecole Polytechnique, IP Paris, F-91128 Palaiseau, France}
\affiliation{$^d$Institute of Metal Physics, 620990, Kovalevskaya str. 18, Ekaterinburg, Russia
}

\date{ \today }

\begin{abstract}
We study the phase diagram and quantum critical region of one of the fundamental models for electronic correlations: the periodic Anderson model. Employing the recently developed  dynamical vertex approximation, we  find 
a phase transition between a zero-temperature antiferromagnetic insulator and a Kondo insulator. In the quantum critical region  we determine a critical exponent $\gamma=2$ for the antiferromagnetic susceptibility. At higher temperatures we have free spins with  $\gamma=1$ instead; whereas at lower temperatures there is an even stronger increase and suppression of the  susceptibility   below and above the quantum critical point, respectively.
\end{abstract}

\maketitle

\let\n=\nu \let\o =\omega \let\s=\sigma

\noindent {\sl Introduction.} Quantum phase transitions are exceedingly exciting since, besides the spatial correlations of a classical phase transition, also (quantum) correlations in time become relevant at zero temperature $T$. This changes the universality class, i.e., the critical exponents, and can be best understood when considering imaginary time $\tau$
which is restricted to $\tau\in[0,1/T]$.  Hence at any finite $T$,
temporal (quantum) correlations are cut off at $1/T$
so that only the spatial correlations remain relevant \cite{Sachdev1999}.

\begin{figure}[tb]
        \centering
                \includegraphics[width=0.45\textwidth,angle=0]{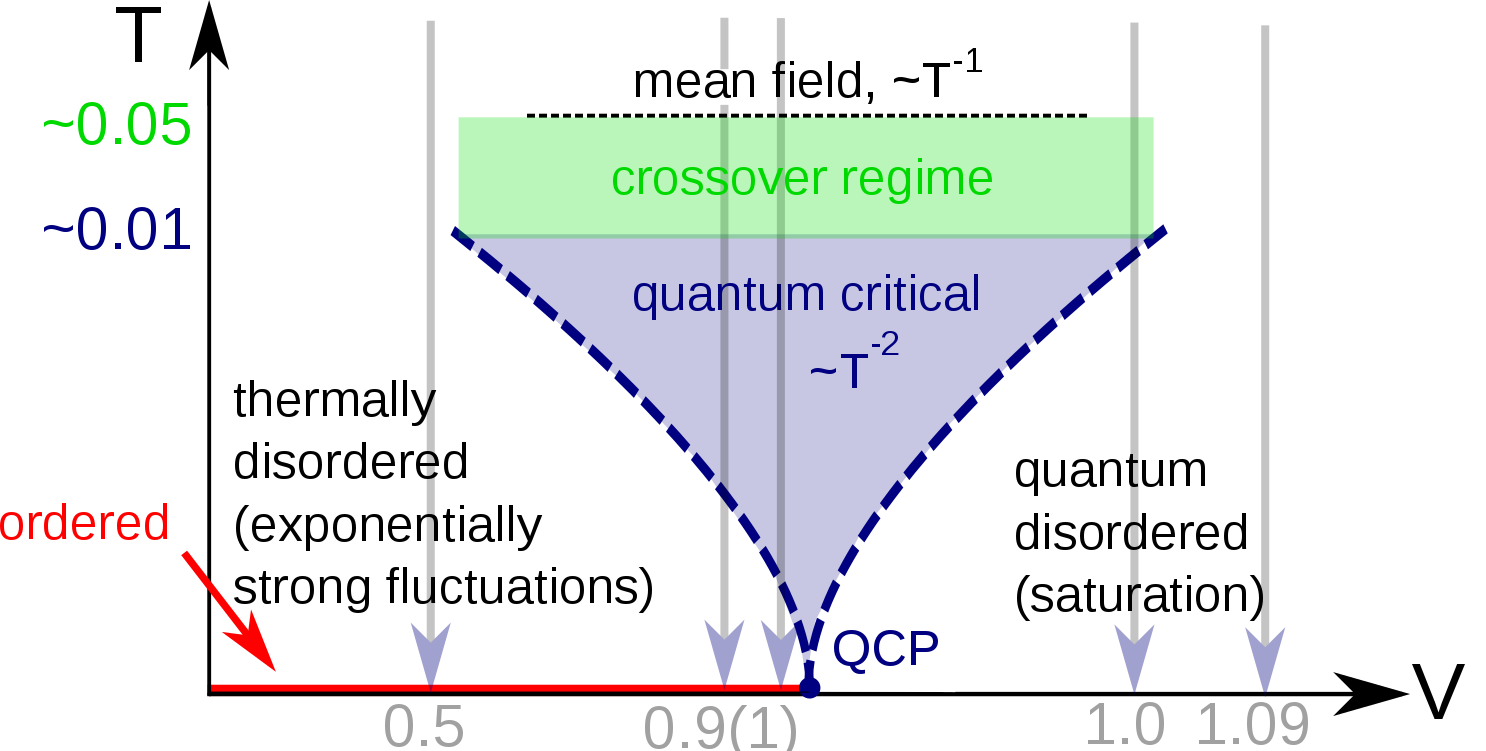}
        \caption{(Color online) \label{Fig1} Schematic phase diagram of the symmetric PAM with a  $T=0$ quantum phase transition towards an antiferromagnetic insulator in $d=2$. Emanating from the  QCP, is a quantum critical region with particular critical exponents. The parameters and values indicate actual D$\Gamma$A results presented below. \label{Fig:Scheme}} 
\end{figure}

Most well studied are, on the experimental side, quantum critical points (QCP's)
in heavy fermion systems \cite{Loehneysen2007, Brando2016} such as  CeCu$_{\text{6-x}}$Au$_{\text{x}}$ \cite{Schroeder2000} and YbRh$_{\text{2}}$Si$_{\text{2}}$ \cite{Custers2003,Paschen2004}. Experimentally accessible is the  unusual behavior within the quantum critical region at a finite $T$ above the QCP; for a schematics see Fig.~\ref{Fig:Scheme}.
The theoretical description of such heavy fermion QCP's is, however, still in its infancy. 

The conventional Hertz\cite{Hertz1976}-Moriya\cite{Moriya1973}-Millis\cite{Millis1993} (HMM)  theory relies on the consideration of the effective $\phi^4$ model for magnetic degrees of freedom and may hence not be applicable for heavy fermion systems with their strong electronic correlations. HMM theory is  by construction a (renormalized) weak-coupling approach which is also valid above the upper critical dimension, i.e., for $d_{\rm eff} =d+z>4$. Here, the spatial dimensions $d$ need to be supplemented by a dynamical exponent $z$, which relates the critical behavior of the correlation length in space ($\xi\sim T^{-\nu}$; $\nu$: critical exponent) and time ($\xi_{\tau}\sim T^{-z\nu}$) at the QCP. Other proposals for a solution of the antiferromagnetic (metallic) criticality problem include the fractionalized electron picture \cite{Sachdev2008}, the critical quasiparticle theory \cite{Senthil2006}, and the strong coupling theory \cite{Abrahams2014}, see also \cite{Vekic1995,Hu2017,Terletska2011,Haldar2016,Lenz2016} for quantum criticality studies employing other  methods.

 Quantum criticality below the upper critical dimension for $d_{\rm eff}=3$ ($d=2$, $z=1$) was considered by Chubukov et al. \cite{Chubukov1994} for the Heisenberg model within a $1/N$ expansion and  by renormalization-group approaches
for Ising symmetry \cite{Sachdev1997,Strack2009}. But again, these approaches cannot be straightforwardly  extended to include fermionic excitations, which are actually essential regarding the experimental realization of QCP's in heavy-fermion systems.  
Despite many promising approaches \cite{Sachdev1999,Si2001,Coleman2005,Kopp2005,Loehneysen2007,StrackThesis}, we hitherto still lack a reliable solution  even for the simplest model for heavy fermion QCP's, the periodic Anderson model (PAM)  beyond a mere (conjectured) mapping onto bosonic models.

In this paper, we hence analyze the QCP of the PAM by means of a recently developed method, the dynamical vertex approximation (D$\Gamma$A) \cite{Toschi2007,Katanin2009}. The D$\Gamma$A  is, similar as related approaches \cite{Rubtsov2008,Rohringer2013,Taranto2014,Ayral2015,Li2015}, a diagrammatic extension of the dynamical mean field theory (DMFT) \cite{Metzner1989,Georges1992,Georges1996}; for a recent review see \cite{RMPVertex}. From the DMFT it inherits a reliable and non-perturbative description of (local) temporal correlations. But on top of these, also non-local spatial correlations are taken into account by means of ladder or parquet diagrams, which do not take the bare interaction but the local irreducible or fully irreducible vertex as a building block. These diagrammatic extensions have been successfully employed for studying critical exponents and phenomena in the Hubbard and Falicov-Kimball model \cite{Rohringer2011,Schaefer2016,Antipov2014,Hirschmeier2015,DelRe2018}. We are hence in the fortunate situation that we can revisit quantum criticality in fermionic models thanks to recent methodological progress.

\noindent
{\sl Model and analytical considerations.} To arrive at a non-mean-field, non-Gaussian critical behavior we  study the PAM in $d=2$ which can be expected to have the same quantum critical exponents as the Heisenberg model, which in turn has a conjectured  $z=1$ \cite{Chubukov1994,Troyer1997}. 
This suggests an effective dimension $d_{\rm eff}=2+1=3$ \footnote{
 Note that  in Ref.~\onlinecite{Schaefer2016} the Hubbard model was studied for $d=3$ which, together with the expected $z=2$ for a metallic antiferromagnetic phase transition, would yield $d_{\rm eff}=5>4$. However, $\nu=1$ instead of the HMM value  $\nu=(d+z-2)/(2z)=3/4$ \cite{Loehneysen2007} was  obtained, because of peculiarities of the Fermi surface, the so-called Kohn lines. This physics, however, can still be understood in terms of Gaussian fluctuations, e.g. in the random phase approximation (RPA) \cite{Schaefer2016}.}.
The Hamiltonian of the PAM reads
\begin{eqnarray}
\mathcal{H} &=& \sum_{{\mathbf k},\sigma} \varepsilon_{\mathbf k} d_{{\mathbf k}\sigma}^{\dagger}d_{{\mathbf k}\sigma}^{\phantom{\dagger}} +  \varepsilon_f \sum_{i\sigma} f_{i\sigma}^{\dagger}f_{i\sigma}^{\phantom{\dagger}}  \nonumber \\ && + U\sum_{i}n_{f,i\uparrow}^{\phantom{\dagger}}n_{f,i\downarrow}^{\phantom{\dagger}} + V \sum_{i,\sigma} \big[ d_{i\sigma}^{\dagger}f_{i\sigma}^{\phantom{\dagger}} +  f_{i\sigma}^{\dagger} d_{i\sigma}^{\phantom{\dagger}}\big]
\label{eq:PAM}
\end{eqnarray}
It consists of localized $f$-electrons with creation (annihilation) operators $f_{i\sigma}^{\dagger}$ ($f_{i\sigma}^{\phantom{\dagger}}$), $n_{f,i\sigma}=f_{i\sigma}^{\dagger} f_{i\sigma}^{\phantom{\dagger}}$, interacting through a local Coulomb repulsion $U$ and  with a local one-particle potential $\varepsilon_f$. Further, there are itinerant $d_{i\sigma}^{\dagger}$ ($d_{i\sigma}^{\phantom{\dagger}}$) electrons with a nearest neighbor hopping $t$, or a corresponding  energy-momentum dispersion relation  $\varepsilon_{\mathbf k}=-2t \left[\cos(k_x)+\cos(k_y)\right]$. Finally, there is a hybridization $V$ between both kinds of electrons.
In the presented calculations, we fix $U\!=\!4t$ (intermediate-to-strong coupling). We consider the half-filled case  $\varepsilon_f =-U/2$, for which the PAM 
maps onto the Kondo lattice model with a coupling 
$J=8 V^2/U$ in the limit $U\gg V$. That is, for large $U$, the $f$-electrons  form localized spins.
This Kondo lattice model shows the famous Doniach \cite{Doniach1977} $T$-$V$ phase diagram, with two competing phases.

On the one hand there is the Kondo effect \cite{Hewson1993}: below the Kondo temperature $T_K$, the spins, that are free at high $T$  with a Curie susceptibility $\chi\sim T^{-1}$,
 get screened. In this case a Kondo resonance forms at the Fermi level. 
In our particle-hole symmetric case of half-filling, this Kondo resonance is however gapped.  This can be understood starting from the non-interacting model ($U=0$): the flat $f$-band at the Fermi energy $E_F$  hybridizes  with the dispersive conduction $d$-band so that  a hybridization gap  opens at   $E_F$. That is, we have a band insulator and for a finite $U$ a quasiparticle-(Kondo-)renormalized picture thereof, i.e. a Kondo insulator. 
For the (single-site) Kondo model 
\begin{equation}
T_K \sim e^{-\frac{1}{\rho_0 J}},
\end{equation}
 where $\rho_0$ is the
non-interacting density of states of the conduction electrons at the Fermi level \cite{Suppl,Hewson1993}. For the PAM we get a similar, somewhat enhanced $T_K$ \cite{Pruschke2000,Suppl}.

Competing with the Kondo effect is a magnetic phase, which can be understood
as the effective Ruderman-Kittel-Kasuya-Yosida (RKKY)  coupling between       $f$-electron spins through the conduction electrons. In second order perturbation theory in $J$, the coupling strength and hence the critical temperature is
\begin{equation}
T_{\rm RKKY} = {\frac{1}{4}}J^2 \chi_{0, \mathbf Q}^{\omega=0},
\label{Eq:RKKY}
\end{equation}
where  $\chi_0$ is the (non-interacting; $V=0$) susceptibility of the conduction electrons {and the factor $1/4=S(S+1)/3$ for spin $S=1/2$ corresponds to the mean-field critical temperature}.
In our case, the maximal coupling appears at the antiferromagnetic (AF) wave vector 
${\mathbf Q}=(\pi,\pi)$. An AF ordering opens a gap, so that we obtain an AF insulator. 
Since $T_K$ is exponentially small for small  $J$ \cite{Doniach1977}, $T_{\rm RKKY}$ prevails for small $J$, whereas at large $J$ the Kondo effect wins. Hence, there is a phase transition from an AF to a Kondo insulator at  $T_K\approx T_{\rm RKKY}$. Hence, the ground state is always insulating. 
At high temperatures, the $f$-electrons are also gapped and form free spins, but the conducting electrons are itinerant; at $T\gtrsim T_K$ the Kondo peak starts to develop but the Kondo insulating gap that is present at lower $T$'s is still smeared out due to strong scattering.

\begin{figure}[tb] 
        \centering
                \includegraphics[width=8.8cm,angle=0]{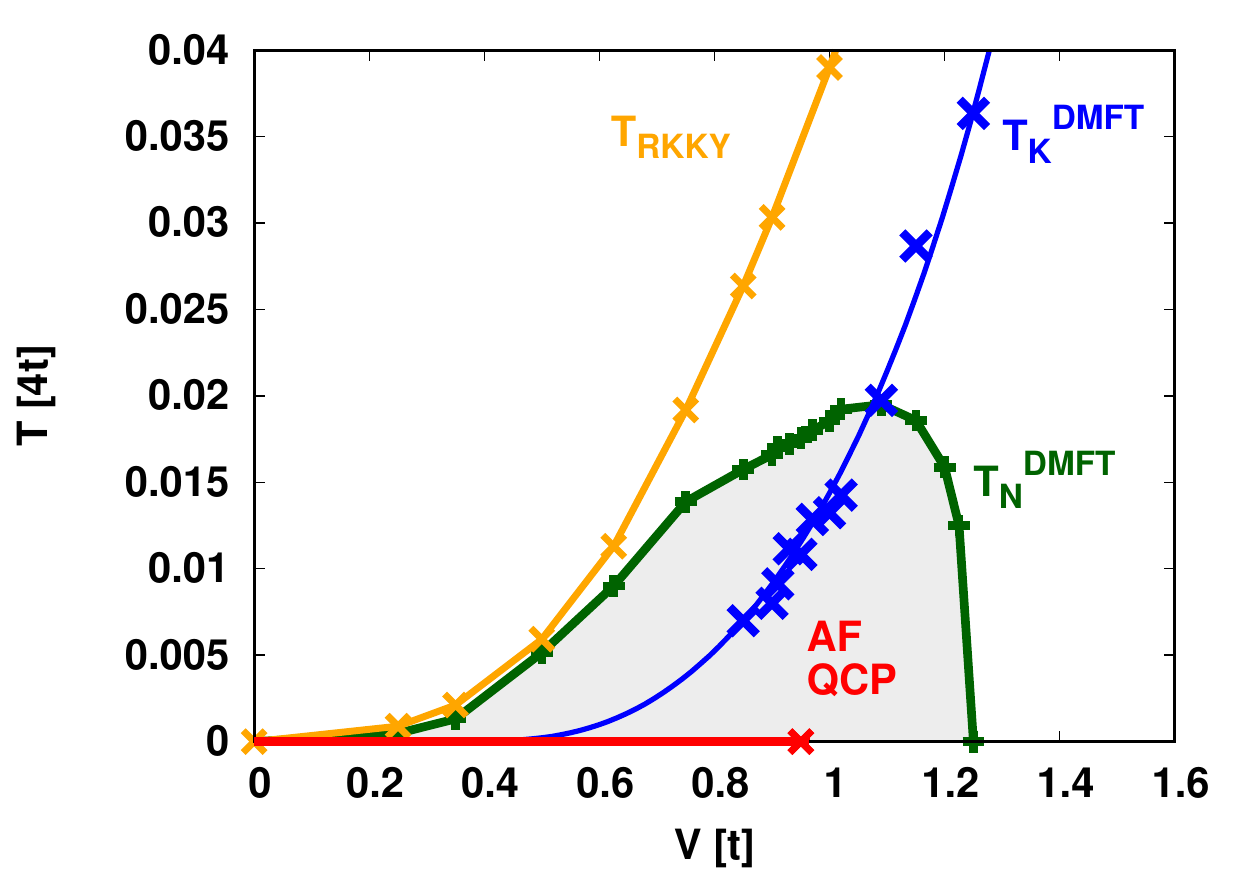}
        \caption{ \label{Fig:Doniach} (Color online) Phase diagram $T$ vs.~$V$ of the 
           half-filled $2d$ PAM  at $U=4t$.
          The figure shows the AF transition $T_N$ line in DMFT (green) and D${\Gamma}$A (red), the DMFT Kondo-temperature $T_{\text{K}}^{\text{DMFT}}$ (blue), and  $T_{\text{RKKY}}$ [yellow, calculated from Eq.~(\ref{Eq:RKKY}), cf. Ref.~7 of the Supplemental Material \cite{Suppl}].}
\end{figure}

\noindent
{\sl Phase diagram.} Fig.~\ref{Fig:Doniach} presents the actual phase diagram of the PAM as calculated using DMFT and D${\rm\Gamma}$A. Here, we employ the ladder D$\Gamma$A with Moriya-$\lambda$ correction \cite{Rohringer2016} which generates spin-fluctuations starting from the local vertex $\Gamma$ calculated for a converged DMFT solution, for further details on the method we refer the reader to \cite{RMPVertex, Rohringer2012, Schaefer2013, Wentzell2016, Kaufmann2017}. 
For the DMFT phase diagram of the Kondo lattice model (and including short-ranged correlations), cf.~\cite{Otsuki2009,Martin2010,Lenz2017}.

\begin{figure*}[ht!]
        \centering
                \includegraphics[width=\textwidth,angle=0]{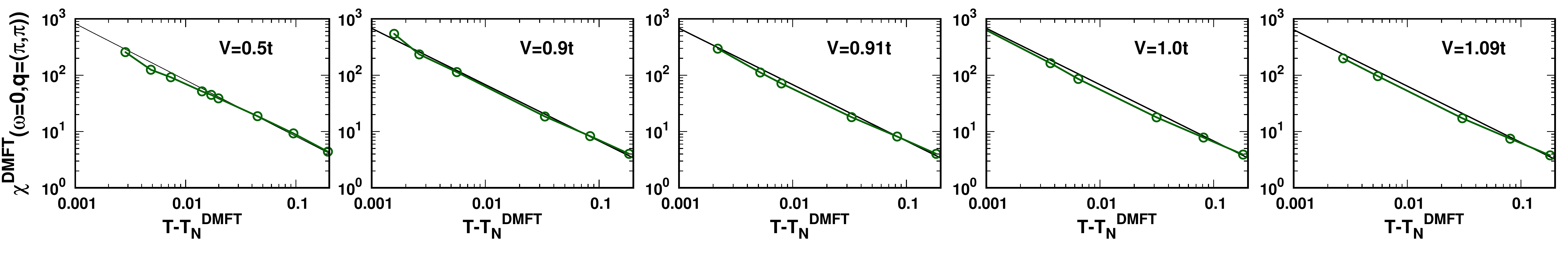}
                \vskip -3.03mm
                \includegraphics[width=\textwidth,angle=0]{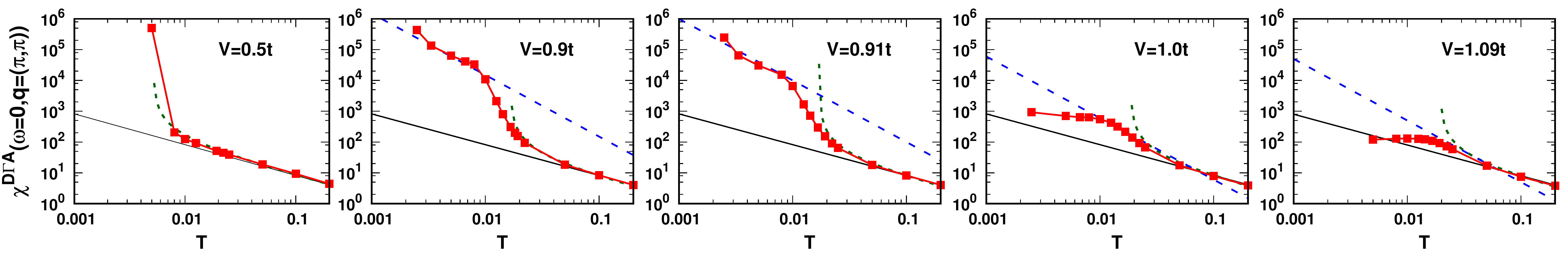}
        \caption{(Color online) \label{Fig3} Magnetic susceptibilities (on a double logarithmic scale)  in DMFT (upper panel, dark green open circles) and D$\Gamma$A (lower panel, red squares).   The black solid and blue dotted lines indicate a $\chi \sim\!T^{-1}$ and $\chi\sim\!T^{-2}$ behavior, respectively{; the green dashed line is the DMFT susceptibility  $\chi \sim\!(T-T_{\rm DMFT})^{-1}$ (black line from the upper panel).} \label{Fig:chi} \label{Fig:susc}}
\end{figure*}

Let us start with the DMFT results, which show AF order at small $V$ in the light-green shaded region of Fig.~\ref{Fig:Doniach}. This order  breaks  down as the Kondo effect sets in and a QCP emerges: there is a $T=0$ phase transition. As we see,  the perturbative result, $T_{\rm RKKY}\sim J^2\sim V^4$ (yellow line), only holds for small $V$; for larger $V$'s DMFT yields a smaller AF transition temperature due to temporal correlations (green line). As we see the AF order breaks down when the DMFT Kondo temperature (blue line, determined from the maximum of the local susceptibility as a function of $T$)
becomes of similar amplitude as the DMFT N\'eel temperature (green line).

The D${\rm\Gamma}$A phase diagram in Fig.~\ref{Fig:Doniach} is distinctively different. Concomitant with the  Mermin-Wagner theorem \cite{Mermin1966}, AF order 
is only found at $T=0$ because of strong non-local fluctuations in $d=2$, 
cf.\ \cite{Katanin2009}
for D${\rm\Gamma}$A  fulfilling the  Mermin-Wagner theorem  for the $2d$ Hubbard model.
 Nonetheless, we have AF order along the red line in Fig.~\ref{Fig:Doniach} and Fig.~\ref{Fig:Scheme}, and hence, at $T=0$, a QCP develops at $V_{\rm QCP}\approx0.91t$.

{\sl Quantum critical region.} Above this QCP region we expect a quantum critical region as visualized in Fig.~\ref{Fig:Scheme}, with non-Gaussian fluctuations. Hence, we study the AF  susceptibility $\chi=\chi^{\omega=0}_{\mathbf Q}$ at momentum ${\mathbf Q}=(\pi,\pi)$ and its critical behavior around the critical $V_{\rm QCP}$ in  Fig.~\ref{Fig:susc}. In DMFT,  $\chi \sim (T-T_N)^{-\gamma} \sim (T-T_N)^{-1}$ see Fig.~\ref{Fig:susc} (upper panels) so that we have a critical exponent $\gamma=1$. 
This reflects the (bosonic) mean-field critical behavior 
of DMFT which neglects spatial fluctuations. At
high temperatures, it smoothly evolves into the  Curie susceptibility
$\chi\sim T^{-1}$ of free spins.

\begin{figure*}[t!]
        \centering
                \includegraphics[width=\textwidth,angle=0]{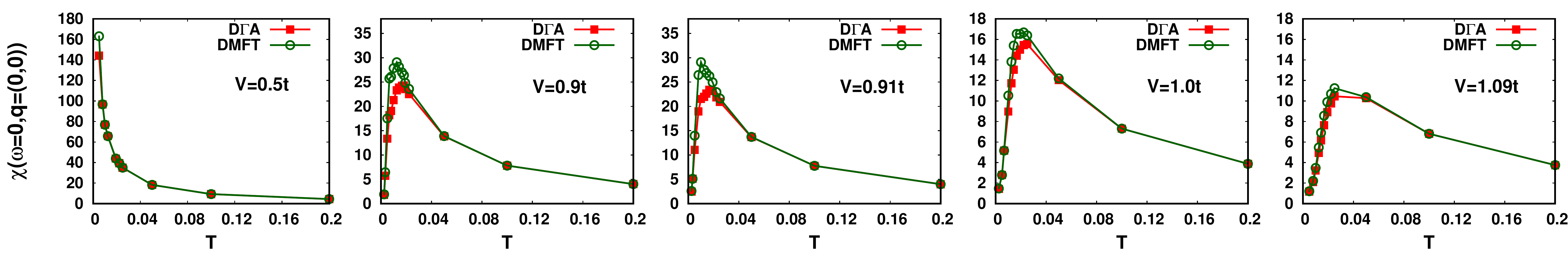}
        \caption{(Color online) Ferromagnetic susceptibility  in DMFT (dark green open circles) and D$\Gamma$A (red squares) for the same parameters as 
in Fig.~\ref{Fig:susc}. \label{Fig:q0susc}}
\end{figure*}

In D$\Gamma$A,  Fig.~\ref{Fig:susc} (lower panels), we observe a completely different behavior. While at high $T$, we have the same $\gamma=1$ Curie behavior, there is a crossover to     $\chi \sim T^{-2}$, i.e., a quantum critical exponent  $\gamma=2$  at lower $T$'s. This critical exponent  and  the related correlation length $\xi \sim T^{-\nu}\sim T^{-1}$ agrees with the 
conjectured  mapping onto a non-linear $\sigma$ model \cite{Chubukov1994,Chakravarty1988}, 
which also displays antiferromagnetic ordering within an insulating phase (as we have) with 
a dynamical critical exponent $z=1$ and yields the same $\xi \sim 1/T$ in the quantum critical regime. 
This yields  the critical exponent  $\nu=1$ for the correlation length, which happens  to be the same critical exponent that one  gets if setting the correlation length in time to its cut-off $\xi_\tau\sim1/T$ and accepting that $z=1$. With the Fisher relation
$\gamma/\nu=2-\eta$  \cite{Fisher1967}, $\gamma\approx 2$ for the susceptibility as observed in Fig.~\ref{Fig:susc} (note that, typically, $\eta$ is vanishingly small even in $d=2$). In the  Supplemental Material \cite{Suppl} Section S.III we present an explanation for this critical exponent on the basis of a sum rule. 

 With increasing dimensionality, we expect the critical exponents at $d\geq 3$ approach their values in HMM theory \cite{Suppl}. Computing quantum critical exponents of strongly correlated electron models such as the PAM was, however, not possible hitherto; quantum Monte Carlo simulations and cluster extensions of DMFT are restricted to too short-ranged correlations.

At the lowest $T$, deviations from this quantum critical behavior are discernible in  Fig.~\ref{Fig:susc} (lower panels)  and are to be expected as we leave the cone-shaped quantum critical region in Fig.~\ref{Fig:Scheme}.
For $V<V_{\rm QCP}$, eventually
antiferromagnetic order sets in at $T=0$.  Already at finite $T$'s, an exponential increase of
the correlation length and the susceptibility with $1/T$ is to be expected  \cite{Chakravarty1988}. A similar exponential scaling was  observed for  the Hubbard model \cite {Schaefer2015-2}. Consistently with this description, one observes a deviation to even larger susceptibilities {at $V\ll V_{\rm QCP}$ and}  lowest $T$'s in Fig.~\ref{Fig:susc}.
 For low $T$ and $V>V_{\rm QCP}$, on the other hand, eventually a Kondo insulating phase develops (quantum disordered phase in Fig.~\ref{Fig:Scheme}).
For this (renormalized) band insulator,  one has $\chi\rightarrow 0$ for $T\rightarrow0$. In agreement with this,  Fig.~\ref{Fig:susc}  shows a deviation to smaller  susceptibilities at  lower  $T$'s; a full suppression of the susceptibility because of the Kondo gap will only occur  at larger $V$ in the accessible $T$-range.

An intriguing, non-universal aspect is the strong enhancement of the susceptibility in the crossover regime between the $\chi\sim1/T$ and  $\chi\sim 1/T^2$ behavior, in particular at $V=0.9$ and $V=0.91$ in Fig.~\ref{Fig:susc}. This originates from enhanced antiferromagnetic correlations,  which  for the periodic Anderson model set in somewhat above $T_{\rm DMFT}$ (see green line in Fig.~\ref{Fig:susc}) and then crossover to the quantum critical $\chi\sim 1/T^2$ region, however with a much larger quantum critical susceptibility (prefactor thereof) 
 than for a Heisenberg model with the exchange interaction providing the same mean-field transition temperature. For a more detailed discussion see the Supplemental Material Section S.4 \cite{Suppl}.


Altogether our results yield the quantum critical region schematically presented in Fig.~\ref{Fig:Scheme}, where we have also inserted the actual $V$ values employed in our calculation, along with the  observed exponents of the $T$-dependence of the susceptibility.

{\sl Uniform susceptibility.} Let us now turn to the ({uniform})
susceptibility, i.e.,  $\chi^{\omega=0}_{\mathbf Q}$ at momentum ${\mathbf Q}=(0,0)$, which has the advantage that it can be  measured more directly in experiment. Its $T$-dependence around $V_{\rm QCP}$ is displayed in Fig.~\ref{Fig:q0susc}.
At large $T$ it shows, similar as the antiferromagnetic $\chi$, the $1/T$ Curie behavior of free spins. However as the spins get screened through the Kondo effect, the ferromagnetic susceptibility shows a maximum around the $T_K$ of Fig.~\ref{Fig:q0susc}, whereas the 
antiferromagnetic susceptibility in  Fig.~\ref{Fig:susc} further grows, signaling the instability toward AF.
Below this maximum, the ferromagnetic susceptibility $\chi^{\omega=0}_{{\mathbf Q}=(0,0)}$ shows essentially in a  $T$-linear behavior in the quantum critical region. Such a behavior has also been reported for a non-linear $\sigma$ model and $1/N$ calculations \cite{Chubukov1994}.



{\sl Conclusion.}  
Thanks to an advanced many-body method, the D$\Gamma$A, we are finally able to
study the phase diagram and even the quantum critical behavior of the PAM, the prime model for heavy fermions,  in $d=2$. We find antiferromagnetic order for small hybridizations $V<V_{\rm QCP}$ at $T=0$, consistent with  the Mermin-Wagner theorem in D$\Gamma$A. In DMFT,  antiferromagnetism breaks down when the Kondo temperature $T_K$ exceeds the N\'eel temperature $T_N$, as in the Doniach scenario, giving rise to a QCP. While  $T_N=0$ in  D$\Gamma$A, we still get 
a comparable $V_{\rm QCP}$, which is 25\% smaller in D$\Gamma$A than in DMFT as the latter neglects non-local  spin fluctuations.  

We identify a quantum critical region with critical exponents $\nu=1$  for the correlation length and $\gamma=2$ for the antiferromagnetic susceptibility, as displayed in Fig.~\ref{Fig:Scheme}; whereas the uniform susceptibility shows a non-critical linear-$T$ dependence. Above the quantum critical region we observe free spins  with $\gamma=1$ at high $T$;  while at small $T$ the  AF susceptibility is  exponentially enhanced in the thermally disordered region $V<V_{\rm QCP}$ and suppressed  in the quantum disordered, Kondo insulating region $V>V_{\rm QCP}$.

Our work  opens a route for studying quantum criticality in various models, which was hitherto only possible for spin models but not for correlated electrons. This removes a blank spot on the map of quantum critical theories, which bears many sophisticated quantum field theoretical considerations, analytical arguments and derivations, but few means to test these numerically in a reliable way.

{\sl Acknowledgments. } We would like to thank Fakher Asaad and Patrick Chalupa for stimulating discussions.  
The present work was supported by the 
European Research Council under the 
European Union's Seventh Framework Program
(FP/2007-2013) through ERC Grant No. 306447,
SFB ViCoM (M.K.,T.S.,K.H.),
Austrian Science Fund (FWF) through the 
Doctoral School ``Building Solids for Function'' (T.S.),  the Erwin-Schr\"odinger Fellowship J 4266 (SuMo, T.S.) and I 2794-N35 (A.T.), as well as  the Russian  Federation  through theme ``Quant'' AAAA-A18-118020190095-4 of FASO (A.A.K.). 
T.S. further acknowledges the European Research Council for the European Union Seventh Framework Program (FP7/2007-2013) with ERC Grant No. 319286 (QMAC) and received funding through the "Exzellenzstipendium Promotio sub auspiciis praesidentis rei publicae" of the Federal Ministry of Education, Science and Research of Austria. Calculations have been done mainly on the Vienna Scientific Cluster (VSC).

\bibliography{main}

\begin{thebibliography}{10}
\bibitem [{Not()}]{NoteSuppl}%
  \BibitemOpen
  \href@noop {} {}\bibinfo {note} {Here, the susceptibility on the r.h.s.\ of
  Eq.~(3) in the main text is calculated at (the self-consistently obtained)
  $T_{\rm RKKY}$, and the onsite term $\chi_{ii}$ has been subtracted as there
  is no self-interaction in the Heisenberg model. We have also accounted for
  that the mean-field coefficient between the N\'eel temperature and the
  exchange parameter $J_H$ is $zS(S+1)/3=1$, as in the considered case $z=4$ is
  the number of nearest neighbors.}\BibitemShut {Stop}%
\end{thebibliography}
\end{document}


\title{Supplemental Material for\\ ``Quantum criticality in the two-dimensional periodic Anderson model''}

\author{T. Sch\"afer$^{a,b,c}$, A. A. Katanin$^d$, M. Kitatani$^a$, A. Toschi$^a$ and K. Held$^a$}

\affiliation{$^a$Institute of Solid State Physics, TU Wien, 1040 Vienna, Austria}
\affiliation{$^b$Coll{\`e}ge de France, 11 place Marcelin Berthelot, 75005 Paris, France}
\affiliation{$^c$CPHT, CNRS, Ecole Polytechnique, IP Paris, F-91128 Palaiseau, France}
\affiliation{$^d$ Institute of Metal Physics, 620990, Kovalevskaya str. 18, Ekaterinburg, Russia
}

\renewcommand{\thefigure}{S.\arabic{figure}}
\renewcommand{\theequation}{S.\arabic{equation}}
\renewcommand{\thesection}{S.\arabic{section}}
\renewcommand{\thetable}{S.\Roman{table}}
\date{ \today }

\begin{abstract}
In this Supplemental material, we first discuss how we extract the DMFT Kondo temperature in Section \ref{Sec:Kondo} before we analyze the effects of finite momentum (and frequency) boxes in 
Section \ref{Sec:grid}. In Section 
\ref{Sec:xi} we present an analytical derivation of the critical exponents for the correlation length at and around the QCP, based on the sum rule Eq.~(\ref{eqn:cond}).  Finally, in Section \ref{Sec:compHeis} we compare to the susceptibility of the Heisenberg model and explain the dramatic increase of the susceptibility in the crossover regime.
\end{abstract}
\maketitle

\section{Determination of the Kondo-temperature in DMFT}
\label{Sec:Kondo}
\begin{figure*}[tb]
        \centering
                \includegraphics[width=\textwidth,angle=0]{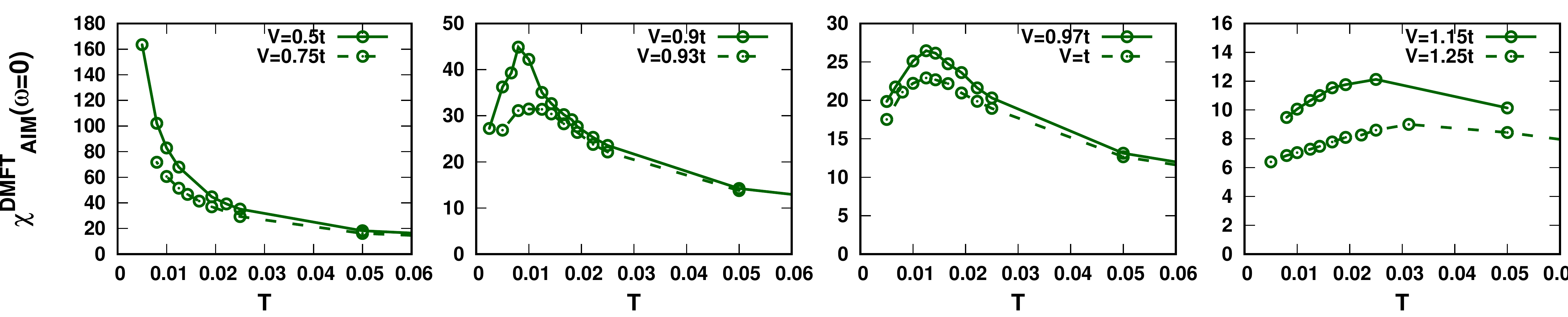}
        \caption{ \label{fig:chis_loc} Temperature dependence of the spin susceptibility of the self-consistently determined AIM in DMFT for $U\!=\!4t$ and several values of the hybridization $V$.}
\end{figure*}

In order to get a reliable estimate of the temperature, where Kondo screening sets in in DMFT, the spin susceptibility $\chi_{\text{AIM}}^{\text{DMFT}}(\omega=0)$ has been calculated from the self-consistently determined Anderson impurity model (AIM) by evaluating the Lehmann representation for the (two-particle) spin susceptibilty.  Fig.~\ref{fig:chis_loc} shows the 
temperature dependence of $\chi_{\text{AIM}}^{\text{DMFT}}(\omega=0)$ at $U\!=\!4t$ and several values of the hybridization $V$.
The Kondo-temperature $T_{K}$ is now determined (where possible, i.e. for $V>0.75t$) as the maximum of $\chi^{\text{DMFT}}_{\text{AIM}}(\omega=0)$. To increase the accuracy of the fit, a polynomial of order 2 is used for interpolating the data around the visible maximum and $T_K$ is determined from this polynomial.

In order to obtain the smooth curve of the phase diagram Fig.~2 (main text), these
$T_K$'s are further fitted to the following function of hybridization V, see Eq.~(6.109) in [\onlinecite{Hewson1993}]:
\begin{equation}
  T_{\text{K}}=AV\exp\left({-\frac{U}{8\rho_{0}V^{2}}+\frac{\pi^{2} \rho_{0} \pi V^{2}}{2 U}}\right),
 \label{eq:fit_kondo}
\end{equation}
with $A$ being the fitted prefactor, $U$ the Coulomb interaction strength, $V$ the hybridization and the non-interacting density of states at the Fermi level is estimated as $\rho_{0}\!\approx\!\frac{1}{W}\!=\!\frac{1}{8t}$. 
For the numerical estimation of $T_{K}$ in the single-site  we refer to [\onlinecite{Krishnamurthy1980,Chalupa2018}], see also [\onlinecite{Kang2019}] for the (doped) periodic Anderson model.

\section{Momentum grid dependence}
\label{Sec:grid}
In the main text, we used 120 momentum grid points for ${\mathbf q}$ in each direction for the interval $[0, \pi]$. Let us here  check the convergence against the number of momentum points. 
In the present calculations, finite grid effects mainly stem from the 
summation $\sum_{\mathbf q} \chi_{\rm m}^{{\rm D}{\Gamma}{\rm A}}({\mathbf q})$ 
to determine the (Moriya) $\lambda$ correction. Here, and in the following
 $\chi_{\rm m}=\chi_{\uparrow\uparrow}-\chi_{\uparrow\downarrow}$ denotes the even spin combination of the susceptibilities as employed commonly in diagrammtic extensions of DMFT, see Ref.~\onlinecite{RMPVertex} where $\chi_{\sigma\sigma'}$ is defined in terms of creation and annihilation operators. The physical susceptibility (in units of the squared Bohr magneton, $\mu_B^2$; ${\mathbf q}$ index suppressed) $\chi=\chi_{\uparrow\uparrow}+\chi_{\downarrow\downarrow}-\chi_{\downarrow\uparrow}-\chi_{\uparrow\downarrow}$ shown in the main paper is a factor of two larger. Note the SU(2) symmetry in the paramagnetic phase, and that the Land\'e factor $g=2$ cancels with  a factor of 1/2 for the spin.

\begin{figure}
\begin{centering}
\includegraphics[width=1.0\columnwidth]{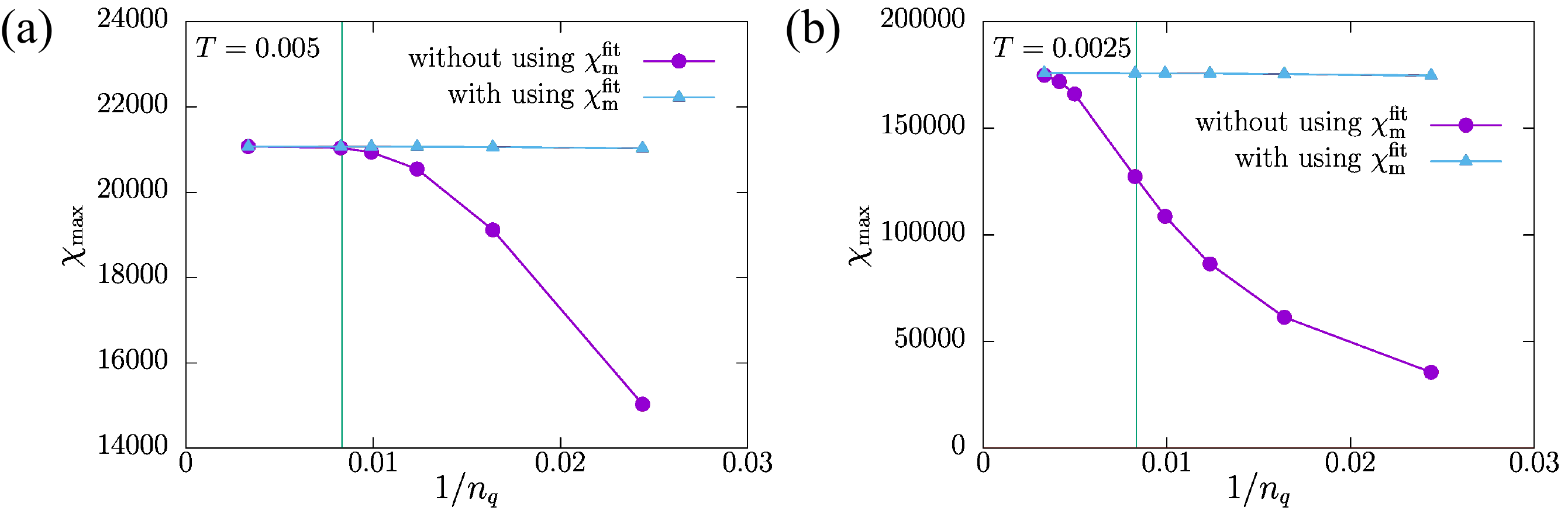}
\par\end{centering}
\caption{Dependence of the maximum magnetic susceptibility $\chi_{\rm m}(\omega=0, {\mathbf q}=(\pi,\pi))$
on $1/n_q$ (momentum grid points in each of the two directions) for (a) $T=0.005$ and (b) $T=0.0025$ (unit: $4t$) with $U=4t$ and $V=0.9t$.
Vertical green lines indicate $n_q=120$ which is used in the main text.
Here, we do not use a frequency extrapolation but use for (a) 120 or for (b) 140 positive frequencies for the vertex.}
\label{fig:LQdep}
\end{figure}

\begin{figure}
\begin{centering}
\includegraphics[width=1.0\columnwidth]{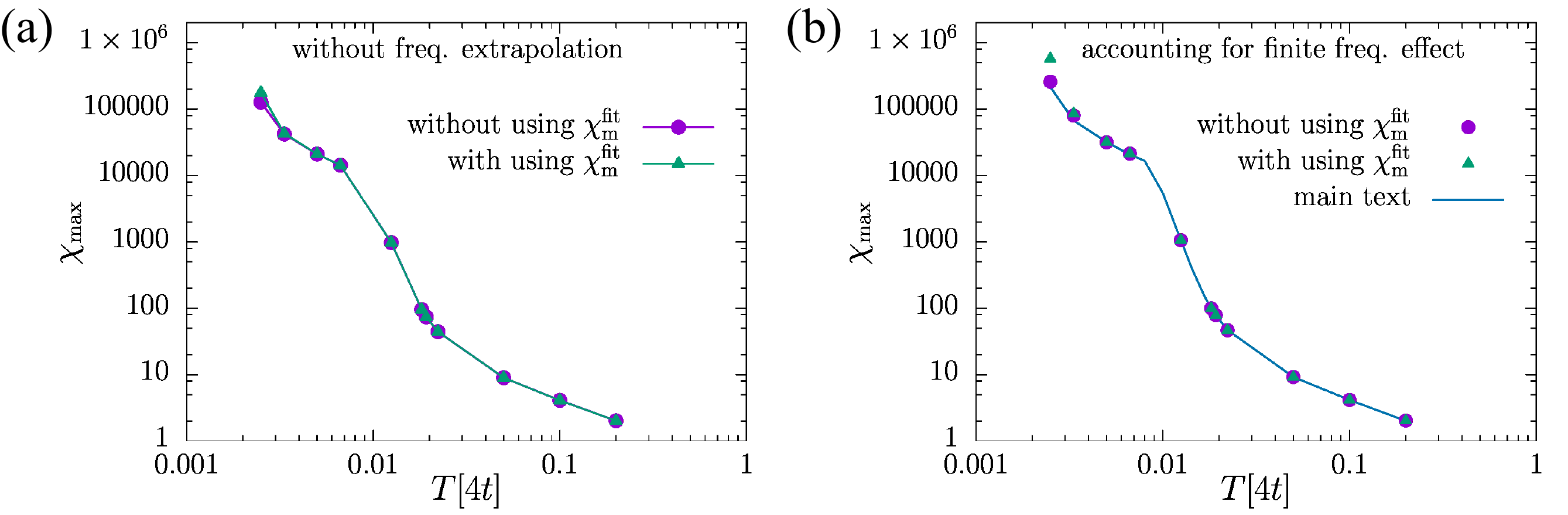}
\par\end{centering}
\caption{Dependence of the maximum magnetic susceptibility $\chi_{\rm m}(\omega=0, {\mathbf q}=(\pi,\pi))$
on $T$ for $U=4t$ and $V=0.9t$ 
(a) without frequency extrapolation and (b) with taking into account the effect of a finite frequency box.  That is
 for (b), instead of using frequency extrapolation as in the main text, 
we consider bare $U$ contribution to the vertex for 1024 positive frequency points as in Ref.~[\onlinecite{Kitatani2018}]. Please note that we obtained almost the same results as the frequency extrapolated ones in the main text (blue line).
}
\label{fig:Tdep}
\end{figure}

We fit $\chi_{\rm m}^{{\rm D}{\Gamma}{\rm A}}({\mathbf q})$ of the form $\chi_{\rm m}^{\rm fit}({\mathbf q})= 
1/(c_1 ({\mathbf q}-{\mathbf Q})^2 +c_2)$ 
with ${\mathbf Q}=(\pi,\pi)$, determining $(c_1,c_2)$ from two points around ${\mathbf Q}$, i.e, 
$\chi_{\rm m}^{{\rm D}{\Gamma}{\rm A}}(\pi,\pi)$ and $\chi_{\rm m}^{{\rm D}{\Gamma}{\rm A}}(\pi,(n_q-2)\pi/(n_q-1))$.
We then use an extremely fine ${\mathbf q}$-grids (we 
analytically integrate for one direction and use $10^6$ grid points for the other direction) 
for the momentum-summation of $\chi_{\rm m}^{\rm fit}$, i.e., we calculate the $\mathbf q$-sum as follows 
\begin{equation}
\sum_q \chi^{{\rm D}\Gamma{\rm A}}_{\rm m}({\mathbf q}) \rightarrow \sum_{\rm q-grid} [\chi^{{\rm D}\Gamma{\rm A}}_{\rm m}({\mathbf q}) -\chi^{\rm fit}_{\rm m}({\mathbf q})]
                                        +\sum_{\rm fine-grids} \chi^{\rm fit}_{\rm m}({\mathbf q}). \label{Eq:qconv}
\end{equation}
Since $(\chi^{{\rm D}\Gamma{\rm A}}-\chi^{\rm fit})$ can be expected to become much smoother  in ${\mathbf q}$-space than $\chi^{{\rm D}\Gamma{\rm A}}$ itself,
a coarser  ${\mathbf q}$-grid should be sufficient for the ${\mathbf q}$-summation.
The results in Fig.~\ref{fig:LQdep} show the dependence of $\chi_{\rm m}$ on the number of $\mathbf q$ points of the ``${\rm q-grid}$'' in Eq.~(\ref{Eq:qconv}) with and without using $\chi_{\rm m}^{\rm fit}$.
While the convergence appears to be much faster when we use $\chi_{\rm m}^{\rm fit}$, 
taking $n_q=120$ without using $\chi^{\rm fit}$ (as in the main text) looks perfectly fine for $T=0.005$ [Fig.~\ref{fig:LQdep}(a)].
For $T=0.0025$ [Fig.~\ref{fig:LQdep}(b)], there is a minute deviation between $n_q=120$ without using $\chi_{\rm m}^{\rm fit}$ and converged results, 
but the temperature dependence plots
in Fig.~\ref{fig:Tdep}(a) and (b) indicate this deviation does not matter for the analysis of the log-log plot in the main text. The same holds for the effect of a finite frequency box, analyzed in  Fig.~\ref{fig:Tdep} (b).  We can hence conclude, that
the momentum and frequency grid  does not affect the scaling of the susceptibilities and, therefore, 
not the conclusions drawn from it.


\section{Magnetic properties and quantum critical exponents in various dimensions $d$}
\label{Sec:xi}

In this Section we discuss the evolution of magnetic properties and quantum
critical exponent for the correlation length $\xi \varpropto T^{-\nu }$ with
dimensionality $d$ in $z=1$ theories. Our consideration is based on the sum rule of the form 
\begin{equation}
T\sum\limits_{i\omega _{n}}\int\limits_{q<\Lambda }\frac{d^{d}q}{(2\pi )^{d}}%
\frac{A}{A^{2}(q^{2}+\xi ^{-2})+\omega _{n}^{2}}=\frac{1}{g},
\label{eqn:cond}
\end{equation}%
where $\omega _{n}$ are the bosonic Matsubara frequencies, $A$, $g$ are some
constants (or weakly temperature dependent quantities). More specifically, $A$ represents the spin-wave (or paramagnon) velocity, while $g$ is the analogue of the dimensionless coupling constant of the non-linear sigma model. Apart
from the non-linear sigma model, describing the Heisenberg model, condition
(\ref{eqn:cond}) is expected to be fulfilled by the Kondo
lattice models, where it represents the sum rule for $f$-electron susceptibility. Performing the summation over Matsubara frequencies and angular integration we obtain   
\begin{equation}
\frac{K_{d}}{2}\int\limits_{0}^{\Lambda }q^{d-1}dq\frac{1}{\sqrt{q^{2}+\xi
^{-2}}}\coth \frac{A\sqrt{q^{2}+\xi ^{-2}}}{2T}=\frac{1}{g},
\end{equation}%
where $K_{d}=1/(2^{d-1}\pi ^{d/2}\Gamma (d/2))$. To treat the $\Lambda $%
-dependence of the integral at large $\Lambda $ we subtract the large
momentum asymptotic of the integrand making the integral convergent and add the result of its analytical integration. This yields 
\begin{equation}
\int\limits_{0}^{\infty }q^{d-1}dq\frac{1}{\sqrt{q^{2}+\xi ^{-2}}}\left[
\coth \frac{A\sqrt{q^{2}+\xi ^{-2}}}{2T}-1\right] +\frac{\Lambda ^{d}\xi }{d}%
\ _{2}F_{1}(\frac{1}{2},\frac{d}{2},1+\frac{d}{2},-(\Lambda \xi )^{2})=\frac{%
2}{gK_{d}},
\end{equation}%
where $_{2}F_{1}(a,b,c,z)$ is the hypergeometric function. Expanding this
function at large negative $z$ we obtain 
\begin{equation}
\left( \frac{T\xi }{A}\right) ^{d-1}f_{d}\left( \frac{T\xi }{A}\right) +%
\frac{1}{d-3}\left[ \ \frac{2\Gamma (d/2)\Gamma (5/2-d/2)}{\sqrt{\pi }(d-1)}-%
\frac{1}{2}(\Lambda \xi )^{d-3}\right] +O((\xi \Lambda )^{d-5})=\frac{2\xi
^{d-1}}{K_{d}}\left( \frac{1}{g}-\frac{1}{g_{c}}\right) ,
\end{equation}%
where 
\begin{equation}
f_{d}(x)=\int\limits_{0}^{\infty }q^{d-1}dq\frac{1}{\sqrt{q^{2}+x^{-2}}}%
\left[ \coth (\sqrt{q^{2}+x^{-2}}/2)-1\right] 
\end{equation}%
and $g_{c}=2(d-1)/(K_{d}\Lambda ^{d-1}).$ Note that $f_{d}(\infty )=2\Gamma
(d-1){\rm Li}_{d-1}(1)$ in $d>2$, where ${\rm Li}_p(x)$ is the polylogarithm function,
${\rm Li}_p(x)=\sum _{k=1}^{\infty } {z^k}/{k^p}$ for integer $p$, and  $f_{2}(x)\simeq 2\ln x$ for $x\rightarrow
\infty $. If we let $\xi \rightarrow \infty $ we obtain $T_{N}\propto \left( 
\frac{1}{g}-\frac{1}{g_{c}}\right) ^{1/(d-1)}$ at $d>2$, while at $d=2$ we
have $T_{N}=0,$ $\xi \varpropto (1/T)\exp[2\pi A(1/g-1/g_{c})/T]$ in
agreement with the results of Ref.\cite{Chubukov1994}.  To obtain the temperature dependence of
the correlation length let us consider the
following cases:

$A.$ $d<3.$ In this case the correlation length solves the equation%
\begin{equation}
\left( \frac{T\xi }{A}\right) ^{d-1}f_{d}\left( \frac{T\xi }{A}\right) =%
\frac{2\xi ^{d-1}}{K_{d}}\left( \frac{1}{g}-\frac{1}{g_{c}}\right) -\ \frac{%
\Gamma (d/2)\Gamma (1/2-d/2)}{2\sqrt{\pi }},
\end{equation}%
(we have transformed $\Gamma$-functions and neglected $\Lambda$-dependent term, which is irrelevant for $d<3$). At the quantum critical point $(g=g_{c})$ we find $\xi \propto 1/T$, 
as was obtained previously for $d=2$ in Ref.~\cite{Chubukov1994}. 

$B.$ $d=3.$ In this case the correlation length solves the equation%
\begin{equation}
\left( \frac{T\xi }{A}\right) ^{2}f_{3}\left( \frac{T\xi }{A}\right) -\frac{1%
}{2}\ln (\Lambda \xi )=\frac{2\xi ^{2}}{K_{d}}\left( \frac{1}{g}-\frac{1}{%
g_{c}}\right) .
\end{equation}%
At the quantum critical point we find $\xi \simeq (3/2)^{1/2}(A/(\pi T))\ln
^{1/2}(A\Lambda /T)$, while above the ordered phase (i.e. at $g<g_c$) outside the classical critical region we obtain $\xi\propto \ln
^{1/2}(A\Lambda /T)/(T^2-T_N^2)^{1/2}$.  

$C.$ $d>3.$ We have 
\begin{equation}
\left( \frac{T}{A}\right) ^{d-1}f_{d}\left( \frac{T\xi }{A}\right) -\frac{1}{%
2(d-3)}(\Lambda ^{d-3}\xi ^{-2})=\frac{2}{K_{d}}\left( \frac{1}{g}-\frac{1}{%
g_{c}}\right) 
\end{equation}
At the quantum critical point we find $\xi \propto T^{(1-d)/2},$ while at $g<g_c$ outside the critical region we obtain $\xi\propto 1/(T^{d-1}-T_N^{d-1})^{1/2}$. 

The obtained 
 results for the temperature dependence of the correlation length at 
$d\geq 3$ agree with the Hertz-Millis-Moriya theory.  

\section{Comparison to the Heisenberg model results}
\label{Sec:compHeis}

The $O(3)$ Heisenberg model result for the static staggered susceptibility in the quantum critical regime reads\cite{Chubukov1994}
\begin{equation}
    \chi_{\bf Q}^{\omega=0}=\frac{ g^2}{\rho_s(1+0.483/N)}\left(\frac{Sc}{\Theta T}\right)^2\approx \frac{ 4}{1.075\rho_s}\left(\frac{Sc}{T}\right)^2 \label{chiQ}
\end{equation}
where
$\Theta=2\ln((5^{1/2}+1)/2)$, $N=3$ is the number of the order parameter components, $g=2$ the Land\'e factor, $\rho_s$ the spin stiffness, and $c$  the spin-wave velocity. Here, we have neglected small corrections due to anomalous exponent $\eta$, which are minute. Within the spin-wave theory,  we obtain the spin stiffness and spin-wave velocity (see, e.g., Refs.~\onlinecite{Irkhin1997,Irkhin1999} and references therein)
\begin{equation}
    \rho_s=J_H S \overline{S}_0 \gamma, \ \ \ \ c=2\sqrt{2} J_H S \gamma \label{Params}
\end{equation}
where $\overline{S}_0\approx 0.3034$ is the ground-state magnetization and $\gamma \approx 1.1571$ is the spin-wave velocity renormalization parameter for $S=1/2$, and $J_H$ is the Heisenberg model exchange parameter. Substituting the results of Eq. (\ref{Params}) into the Eq. (\ref{chiQ}) we find for $S=1/2$
\begin{equation}
    \chi_{\bf Q}^{\omega=0}\approx 
     14\frac{ J_H}{{T}^2} \label{chiQ1}
\end{equation}
We note that the result of Eq. (\ref{chiQ1}) is applicable at $T\ll J_H$. 

\begin{figure}[tb]
        \centering
                \includegraphics[width=0.4\textwidth,angle=0]{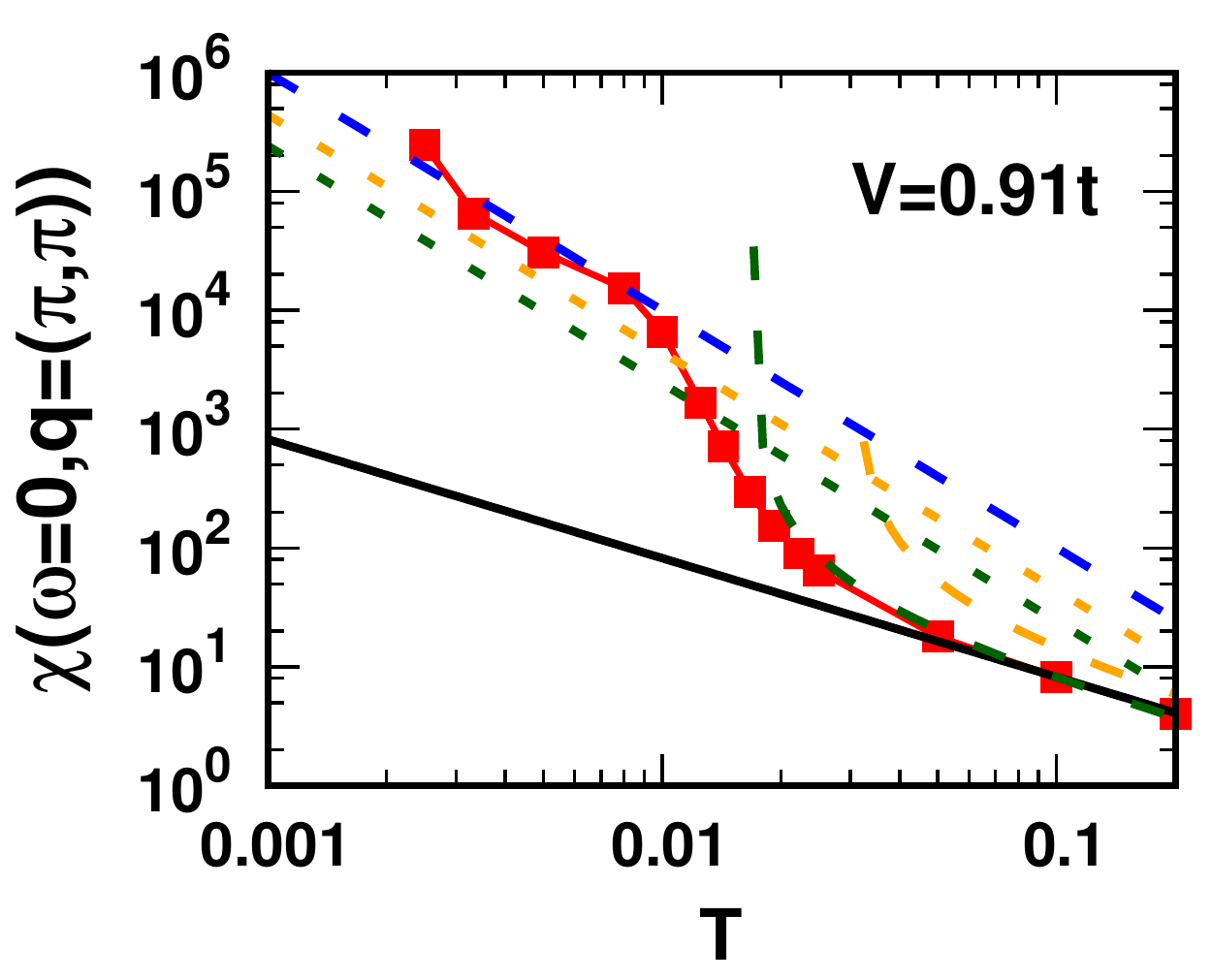}
        \caption{(Color online) \label{Fig:V091} Antiferromagnetic susceptibility vs. temperature of the periodic Anderson model at $V=0.91$ in  D$\Gamma$A (red squares and line) compared to a free spin $1/T$ (black line) and quantum critical $J^*/T^2$ behavior (dashed blue line). The deviation from the $1/T$ behavior sets in when also the DMFT susceptibility  [green long-dashed  line; $\chi \sim 1/(T-T_{\rm DMFT})$] starts to deviate from $1/T$, but eventually non-local fluctuations lead to a lower susceptibility and the dashed blue quantum critical behavior.
The RKKY mean field behavior [$\chi \sim 1/(T-T_{\rm RKKY})$; yellow long-dashed line]
would instead suggest a deviation from the free spin $1/T$ behavior at higher temperatures. The obtained quantum critical behavior is compared to the quantum critical behavior Eq.~(\ref{chiQ1}) expected for the Heisenberg model, with $J_H=T_{\rm RKKY}$ (yellow short-dashed  line) and  $J_H=T_N^{\text{DMFT}}$ (green  short-dashed line).} 
        \label{FigHeis}
\end{figure}

For small hybridization $V$, the exchange parameter of the Heisenberg model will be given by RKKY, i.e.,\cite{NoteSuppl} $J_H=T_{\rm RKKY}$.
This curve is shown as the yellow short-dashed line in  Fig.~\ref{FigHeis}. For the green short-dashed line we instead employ the DMFT transition temperature, i.e., $J_H=T_{N}^{\rm DMFT}$. 

Let us now compare this quantum critical behavior (\ref{chiQ1}) to that obtained in D$\Gamma$A for the  periodic Anderson model (red dots  in Fig.~\ref{FigHeis}). One can see that despite using a not too large $U=4t$ and not too small $V=0.91t$ in the periodic Anderson model, the obtained susceptibility in the quantum critical region  (blue line in Fig. \ref{FigHeis}) qualitatively agrees with the result of the Heisenberg model (\ref{chiQ1}). But the susceptibility for the periodic Anderson model is even slightly larger (has a larger fitted prefactor $J^*$) than for the effective Heisenberg model with  $J_H=T_{\rm RKKY}$. That is, we find  for the blue dashed line in Fig.~\ref{FigHeis}
\begin{equation}
    \chi_{\bf Q}^{\omega=0}= \frac{ J^*}{{T}^2} \approx 
     20\frac{ J_H}{{T}^2} \; ; \;\;\; \mbox{ where } J_H=T_{\rm RKKY} \;,
\end{equation}
i.e., $J^*>14 \, T_{\rm RKKY}$  [Eq.~(\ref{chiQ1}) with $J_H=T_{\rm RKKY}$]. 
At the same time,  the crossover temperature  $T^*=T_{\rm DMFT}$ is considerably smaller than in the Heisenberg model with $T^*=J_H=T_{\rm RKKY}$. That is, we have a much larger ratio $J^*/T^*$, and consequently a more dramatic increase of the susceptibility  in the crossover regime. 

This can be  visualized by means of  Fig.~\ref{FigHeis}: For the periodic Anderson model the susceptibility is first enhanced compared to the  free spin behavior (black in Fig.~\ref{FigHeis}) in a mean field way $\sim 1/(T-T_{\rm DMFT})$ at $T\gtrsim T_{\rm DMFT}$ (green long-dashed line  in Fig.~\ref{FigHeis}), before turning to the quantum critical behavior (blue dashed line).
For the Heisenberg model with $J_H=T_{\rm RKKY}$, we would have instead a crossover from the black line of free spins in  Fig. \ref{FigHeis} at high temperatures, to the yellow short-dashed  RKKY mean-field $\chi \sim 1/(T-T_{\rm RKKY})$  at the beginning of the crossover regime to the yellow long-dashed in the quantum critical regime.

The reason for this difference between the Heisenberg model with  $J_H=T_{\rm RKKY}$ and the periodic Anderson model is twofold:
(i) There are terms beyond second order in $J$ ($J$: Kondo coupling) which not only modify $J_H$ but also lead to additional multiple spin interactions beyond the Heisenberg model.
(ii) We are in a region with Kondo screening. Hence corrections to a mere spin model are to be expected.

\bibliography{main}